\documentclass[aps,prb,reprint,longbibliography,superscriptaddress]{revtex4-2}

\usepackage{amsmath}
\usepackage{amssymb}
\usepackage{xcolor}

\usepackage{amsmath,amssymb,amsfonts,bm,color,graphicx}

\usepackage{tabularx,multirow,array,diagbox}

\usepackage{bm}

\usepackage[unicode=true,colorlinks=true]{hyperref}

\usepackage[etex=true,export]{adjustbox}

\usepackage{ulem}
\renewcommand{\v}[1]{{\bf #1}}

\newcommand{\beq}{\begin{equation}}
\newcommand{\eeq}{\end{equation}}
\newcommand{\beqn}{\begin{eqnarray}}
\newcommand{\eeqn}{\end{eqnarray}}

\usepackage{graphicx} 

\begin{document}

\title{Edge dependent Josephson Diode effect in WTe$_{2}$-Based Josephson junction}

\author{Guo-Liang Guo}
\email{sjtu2459870@sjtu.edu.cn}
\affiliation{Tsung-Dao Lee Institute, Shanghai Jiao Tong University, Shanghai 201210, China and School of Physics and Astronomy, Shanghai Jiao Tong University, Shanghai 200240, China}
 
\author{Xiao-Hong Pan}
\affiliation{Department of Physics, College of Physics, Optoelectronic Engineering, Jinan University, Guangzhou 510632, China}

\author{Hao Dong}
\affiliation{School of Physics and Wuhan National High Magnetic Field Center, Huazhong University of Science and Technology, Wuhan, Hubei 430074, China}

\author{Xin Liu}
\email{phyliuxin@hust.edu.cn}
\affiliation{Tsung-Dao Lee Institute, Shanghai Jiao Tong University, Shanghai 201210, China and
School of Physics and Astronomy, Shanghai Jiao Tong University, Shanghai 200240, China}
\affiliation{School of Physics and Wuhan National High Magnetic Field Center, Huazhong University of Science and Technology, Wuhan, Hubei 430074, China}
\affiliation{Hefei National Laboratory, Hefei 230088, China National Laboratory, Hefei 230088, China}
\affiliation{Shanghai Research Center for Quantum Sciences, Shanghai 201315, China}

\begin{abstract}
The Josephson diode effect (JDE), a nonreciprocal supercurrent, is a cornerstone for future dissipationless electronics, yet achieving high efficiency in a simple device architecture remains a significant challenge. Here, we theoretically investigate the JDE in a junction based on monolayer 1T'-WTe$_2$. We first establish that different edge terminations of a WTe$_2$ nanoribbon lead to diverse electronic band structures, some of which host asymmetric edge states even with crystallographically equivalent terminations. This intrinsic asymmetry provides a natural platform for realizing the JDE. With a WTe$_2$-based Josephson junction, we demonstrate a significant JDE arising purely from these asymmetric edges when time-reversal symmetry is broken by a magnetic flux.  While the efficiency of this edge-state-driven JDE is inherently limited, we discover a crucial mechanism for its enhancement: by tuning the chemical potential into the bulk bands, the interplay between edge and bulk transport channels boosts the maximum diode efficiency more than $50\%$. Furthermore, we show that this enhanced JDE is robust against moderate edge disorder. Our findings not only propose a novel route to achieve a highly efficient JDE using intrinsic material properties but also highlight the potential of engineered WTe$_2$ systems for developing advanced superconducting quantum devices.
\end{abstract}

\maketitle

\section{introduction}

The semiconductor diode, whose foundational operating principles stem from the seminal p-n junction \cite{Scaff1947, Shockley1949}, constitutes an indispensable class of electronic components with ubiquitous technological implementations. A conceptually analogous yet physically distinct phenomenon emerges in superconductors (SC), a phenomenon known as the superconducting diode effect (SDE), which allows the superconducting current to flow only in one direction \cite{Jiang2022}. This nonreciprocal supercurrent phenomenon has garnered substantial experimental and theoretical attention in contemporary condensed matter physics \cite{Jiang1994, Braginski2019,Ando2020,Miyasaka2021,Zhang2022,He2022,Tanaka2022,Bauriedl2022,Liang2023,Picoli2023,Satchell2023,Hou2023,Banerjee2024,Debnath2024}, due to its revelation of novel transport mechanisms in symmetry-broken superconducting systems, and its potential for unprecedented device integration prospects. This asymmetric supercurrent rectification differs fundamentally from conventional semiconductor diode operation, enabling innovative dissipationless electronics that could potentially revolutionize superconducting circuit design. In superconducting systems, Josephson junctions and superconducting quantum interference devices (SQUIDs) are the typical superconducting devices \cite{Yu2002,Ioffe1999}. The nonreciprocal supercurrent effect in Josephson junctions is then termed the Josephson diode effect (JDE) \cite{Chen2018,Baumgartner2022,Davydova2022,Pal2022,Steiner2023,Banerjee2023,Legg2023,Maiani2023}, which has different magnitudes of critical currents ($|I_{c+}|\neq |I_{c-}|$) in opposite directions. This has been experimentally realized and theoretically calculated in various systems with broken time-reversal and inversion symmetries \cite{Satchell2023,Hou2023,Banerjee2024}. The time-reversal symmetry breaking is usually achieved by applying an external magnetic field \cite{Yokoyama2014,Alidoust2021,Yuan2022,Legg2022,Jeon2022,Lu2023}, while the inversion symmetry breaking can arise from the asymmetry of stacked heterostructures \cite{Ando2020,Baumgartner2022}, artificial device configuration \cite{Turini2022,He2023}, interface \cite{Guo2024}, and so on \cite{Bauriedl2022,Yun2023}. Furthermore, the field-free JDE has also been explored in various systems where the time-reversal symmetry is spontaneously broken by valley polarization \cite{Lin2022,DiezMerida2023,Hu2023}, (anti)ferromagnetic materials \cite{Gao2024,Zhang2025}, loop currents \cite{Qi2025}, and exotic pairing function induced in unconventional superconductors \cite{Zhu2023,Ghosh2024}. Moreover, the JDE can also be realized in asymmetric supercurrent interferometers or SQUID-like devices \cite{Souto2022,Haenel2022,Paolucci2023,Li2024}. However, the diode efficiency $\eta=(I_{c+}-|I_{c-})|/({I_{c+}+|I_{c-}|})$ is usually low \cite{Ando2020,Davydova2022,Guo2024} or the device is complex with more than one Josephson junction to gain the large diode efficiency \cite{Souto2022,Haenel2022}. On the other hand, the emergent discovery and technological implementation of novel material systems exhibiting exotic quantum phenomena has become a hallmark of modern condensed matter physics, with paradigmatic examples including (anti)ferromagnetism \cite{Gingrich2016,Szombati2016}, altermagnetism \cite{Krempasky2024,Liu2024}, and topological surface states and edge states \cite{Fu2009,Badiane2011,Cho2013,Oostinga2013}, as well as boundary sublattice physics \cite{Kheirkhah2022, Zhu2023a}, etc. Additionally, transition metal dichalcogenides (TMDs) are at the forefront of condensed matter research, as they host a rich variety of intriguing ground states, such as topological insulators (TI) and semimetals \cite{Qian2014,Deng2016}, charge density waves \cite{Ritschel2015,Li2016}, and various types of possibly unconventional superconductivity \cite{Yuan2014,Lu2015,Xi2016}. This fundamental progress is predicated upon systematic investigations that establish viable pathways toward the realization of JDE and robust property optimization (including key performance metrics such as diode efficiency) in engineered quantum systems.

In this work, we propose to implement the JDE in 1T'-WTe$_2$ based Josephson junction utilizing external magnetic flux. The 2D 1T'-WTe$_2$ functions as a quantum spin Hall material characterized by strong crystalline anisotropy. With 1T'-WTe$_2$ based Josephson junction, we demonstrate distinct edge-state ensembles due to varying edge-site atomic configurations (Fig.~\ref{dev}(a)), thereby enabling JDE through asymmetric SQUID-like geometric configurations in the presence of magnetic flux. The polarity of diode efficiency is contingent upon the magnitude of the magnetic flux and edge configurations. Crucially, we found that the maximal magnitude of diode efficiency can be enhanced to more than 53\% when the fermi surface is tuned to the bulk with the asymmetric edge states maintained (Fig.~\ref{dev}(c)). Finally, we find that edge disorder in the system has little effect on the large JDE. Our work provides a new material platform to achieve JDE with large diode efficiency and demonstrates a great potential of future applications.

\section{Termination-Dependent dispersion in the WTe$_{2}$}
    

\begin{figure}[!htbp]
	\centering
	\includegraphics[width=1\columnwidth]{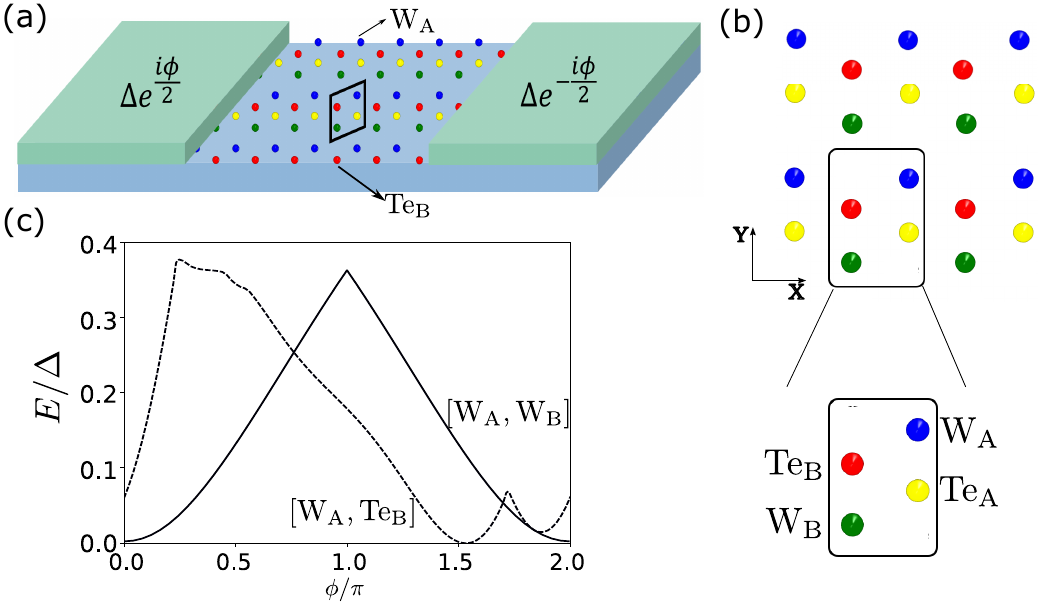}
	\caption{(a) The setup of 1T' WTe$_2$ based Josephson junction with the termination $\mathrm{[W_A, Te_B]}$. (b) Lattice structure and unit cell of monolayer WTe$_2$. The colored circles indicate Wannier functions contributing to the low-energy physics. The other dashed circles in the unit cell do not play a role in the effective low-energy electronic structure. (c) The Josephson potential with the maximal diode efficiency (dashed line) with the inequivalent edges, and the Josephson potential with no diode effect (solid line) with the equivalent edges.
  }
\label{dev}
\end{figure}
WTe$_2$ is one of the transition metal dichalcogenide (TMDC) materials known for its variety of polytypic structures, 2H, 1T, and 1T' \cite{Qian2014}. The 2H structure, identified as a trivial semiconductor \cite{Qian2014}, consists of three planes of hexagonally packed atoms in two dimensions, Te-W-Te, arranged in an ABA stacking pattern. Conversely, the 1T structure entails three planes with ABC stacking; however, it is unstable \cite{Qian2014}, and undergoes lattice distortion along the x-direction, resulting in the stable 1T' structure with a rectangular unit cell (Fig.~\ref{dev}(b)). This distortion renders the lattice nonsymmorphic with the glide-mirror $M_x$ and twofold rotation symmetry \cite{Ok2019,Hsu2020,Nandy2022}, each accompanied by a half lattice translation in the x-direction (Fig.~\ref{dev}(b)). Additionally, the unit cell structure respects time-reversal and inversion symmetries. The monolayer 1T' configuration of WTe$_2$ is known as the topological nontrivial phase with a band inversion at the $\Gamma$ point \cite{Choe2016,Muechler2016}.

To investigate the JDE in the 1T' WTe$_2$ based Josephson junction, we begin with the tight-binding (TB) model based on DFT calculations \cite{Ok2019}. The four bands closest to the Fermi level originate from the two $d_{x^2-y^2}$-type orbitals of the W atoms and two $p_x$-type orbitals of the Te atoms. The Hamiltonian, based on the spin and four Wannier orbitals, is expressed as:
\begin{equation}
    H=[h_0(k)-\mu]\otimes s_0 + V_{soc}\rho_z\tau_ys_z,
    \label{ham}
\end{equation}
where $\rho,\tau,s$ label the $\mathrm{(W,Te)_i(i=A,B)}$ sublattices, Wannier orbitals $p,d$ and spin degree of freedom. 

\begin{figure}[!htbp]
	\centering
	\includegraphics[width=1\columnwidth]{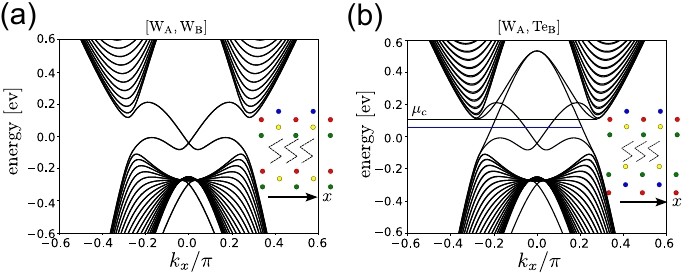}
	\caption{(a)(b) Band structure with the periodic condition in the x-direction, open boundary condition in the y-direction with the width of $W=11$. (a) The band structure with the termination $\mathrm{[W_A, W_B]}$. (b) The band structure with the termination $\mathrm{[W_A, Te_B]}$, the blue line is the chemical potential in the bulk and used in Fig.~\ref{edge-JDE}, $\mu_c$ the position of the bottom of the conduction band. 
  }
\label{band}
\end{figure}

Utilizing the TB Hamiltonian, we proceed to analyze the band structure under various terminations. We construct a lattice model with periodic boundary conditions in the $x$-direction and open boundary conditions in the $y$-direction with a width $2W$. Accordingly, both the upper and lower edges of the slab in the $x$ direction feature four distinct terminations denoted as $\mathrm{W(Te)_i}(i=A, B)$. We represent the terminating atoms of the upper and lower edges using the set $[\mathrm{W(Te)_i, W(Te)_i}]$. Theoretically, there are $4\times4=16$ potential configurations in total. Nevertheless, due to the inversion symmetry of the unit cell, $\mathrm{W(Te)}_A$ is equivalent to $\mathrm{W(Te)}_B$, resulting in redundant configurations, such as $[\mathrm{W_A, W_A}]$ = $[\mathrm{W_B, W_B}]$, reducing the count to ten distinct cases (Appendix A). Fig.~\ref{band}(a) illustrates a representative band structure with termination $[\mathrm{W_A, W_B}]$ (inset in Fig.~\ref{band}), which features degenerate edge states. This degeneracy arises from the inversion symmetry connecting the upper $\mathrm{W_A}$ atoms and the lower $\mathrm{W_B}$ atoms. In contrast, the band structure associated with the termination $[\mathrm{W_A, Te_B}]$ (Fig.~\ref{band}(b)) displays non-degenerate edge states, where each edge exhibits different Fermi velocities owing to the non-equivalent terminating atoms. Additional band structures for various termination configurations are detailed in Appendix A.

The scenarios vary when periodic boundary conditions are imposed in the $y$ direction and open boundary conditions in the $x$ direction with a length of $2L$. Within this framework, two distinct terminations, $\mathrm{W_A Te_A}$, and $\mathrm{W_B Te_B}$ exist for the two edges of the slab along the $y$ direction. We represent these edge configurations with $[\mathrm{(WTe)_i, (WTe)_i}]$. In principle, there are $2\times2=4$ possible configurations of the slab in total. Remarkably, all four configurations are equivalent due to mirror symmetry $M_x$, which relates the two edges in the $[\mathrm{(WTe)_i, (WTe)_i}]$ termination, and inversion symmetry, which connects the edges in the $[\mathrm{(WTe)_i, (WTe)_{\bar{i}}}]$ termination, with $\bar{i}$ denoting the opposite of $i$. The band structures display identical characteristics, featuring degenerate edge states with a Dirac point crossing at $k_y=\pi$ \cite{Ok2019} (Appendix A). Consequently, to implement the necessary asymmetry for the JDE, the Josephson junction must be constructed along the $x$ direction.




\section{JDE with edge distortion}
We consider the JDE in WTe$_2$ based Josephson junction, which is composed of a 2D WTe$_2$ covered with superconductors on both ends. Given that the band structure along the y-direction presents symmetric edges due to inversion and mirror symmetries (Appendix A), we thus focus on the junction along the x-direction, as shown in Fig.~\ref{dev}(a). Without loss of generality, we here consider the edges formed by $\mathrm{W_A}$ and $\mathrm{Te_B}$ atoms,  with the associated band dispersion depicted in Fig.~\ref{band}(b). Then, we construct a TB Bogoliubov–de Gennes (BdG) Hamiltonian of SC/TI/SC junction with the WTe$_2$ Hamiltonian Eq.~\eqref{ham} and compute the Andreev levels of the system by calculating the eigenenergies of the TB Hamiltonian as a function of the superconducting phase difference $\phi$. The magnetic field is applied via the Peierls substitution \cite{Datta1995}, the hopping parameters change to $t\to e^{-ie/\hbar\int \mathbf{A}\cdot dl}$. With the Landau gauge $\mathbf{A}=(By,0,0)$, the phase $e/\hbar\int \mathbf{A}\cdot dl=\pi y\frac{Ba^2}{\phi_0}$, with $y$ and $a$ the $y$-direction coordinate and lattice constant, respectively. Notably, a $4\pi$ Josephson effect is anticipated in the ideal case with conserved fermion parity within a topological insulator-based Josephson junction. However, quasiparticle poisoning can induce fermion parity switches, leading to a $2\pi$ Josephson effect \cite{Chen2018,Pribiag2015}, and we consider this scenario in our analysis. 


\begin{figure}[!htbp]
	\centering
	\includegraphics[width=1\columnwidth]{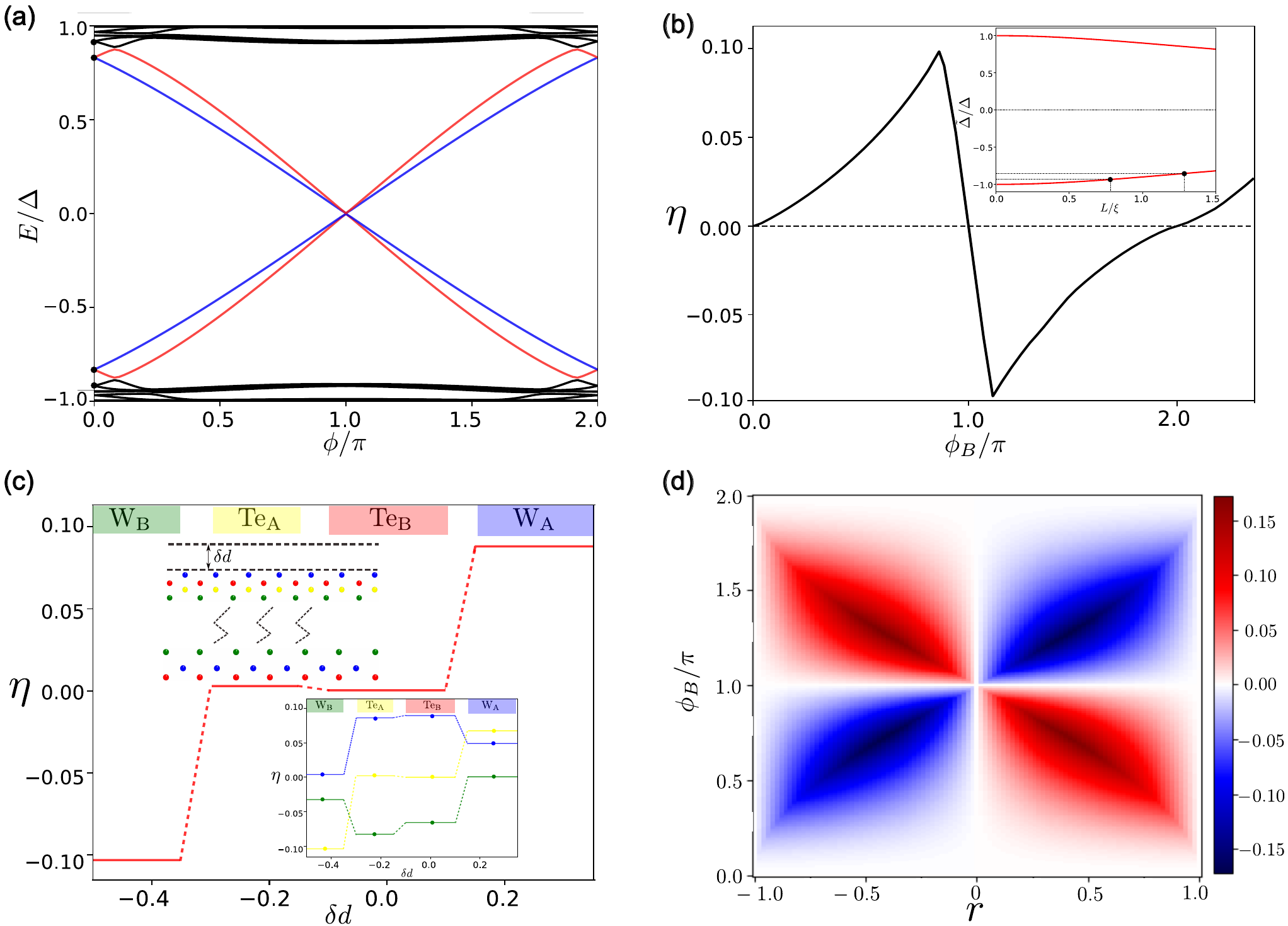}
	\caption{(a) Andreev levels of the junction for $L=20a$, with $L/\xi_{u(d)}\approx0.83,1.33$ for the two edge states, the two colors correspond to two edge states. The black dots at $\phi=0$ correspond to the black dots in the inset in (b). (b) Diode efficiency changes with magnetic flux; the inset shows the Eigenenergies at $\phi=0$ ($\tilde{\Delta}/\Delta$) change with junction length. (c) Diode efficiency as a function of the upper edge configuration while fixing the lower boundary atoms $\mathrm{Te_{B}}$, at a constant magnetic field $\phi_B/\pi=0.86$. The upper inset schematically shows the system. The lower inset shows the diode efficiency as a function of the upper edge configuration while fixing the lower boundary atoms $\mathrm{W_A, W_B, Te_{A}}$. The lines' color corresponds to four distinct lower boundary atoms, the colored regions indicate different upper boundary atom configurations (displayed above), with colors matching Fig.~\ref{dev}. (d) Diode efficiency changes with asymmetric $r$ and magnetic flux $\phi_B$ with the toy potential Eq.~\eqref{edge-pot}.
  }
\label{edge-JDE}
\end{figure}

When the Fermi level falls within the bulk gap (indicated by the blue line in Fig.~\ref{band}(b)), only two edge states can facilitate Cooper pair tunneling.
The mismatch in Fermi velocities between the two edge states leads to differing coherence lengths for these states, $\xi_u\neq\xi_d$. As a result, the Andreev level originating from the top and bottom edges differs, as shown in Fig.~\ref{edge-JDE}(a), effectively acting as two inequivalent Josephson junctions in a SQUID-like configuration. Furthermore, due to time-reversal and particle-hole symmetries, the Andreev level energies approach zero when the phase difference $\phi=\pi$. The associated edge junction Andreev levels in the short junction limit ($L/\xi\ll1$) take the form \cite{Beenakker1991}
\begin{equation}
    E(\phi)=\pm\tilde{\Delta}\sqrt{1-T\sin^2\left(\frac{\phi}{2}\right)},
\end{equation}
with $T$ the transmission amplitude and $T\to1$ for TI edge states \cite{Koenig2008,Guo2024a}. $\tilde{\Delta}$ represents the eigenenergies of the junction at $\phi=0$, which captures the junction length $L$ and coherence length $\xi$. This leads to the simplified form of the Andreev levels as $E(\phi)=\pm\tilde{\Delta}|\cos(\phi/2)|$. With the perfect Andreev reflection in TI-based Josephson junctions \cite{Koenig2008}, the dependence of $\tilde{\Delta}$ on the junction length and coherence length is \cite{Schaepers2001}
\begin{equation}
    \left(\frac{\tilde{\Delta}}{\Delta}\right)\left(\frac{L}{\xi}\right)=2\arccos\left(\frac{\tilde{\Delta}}{\Delta}\right),
\end{equation}
with $\Delta$ the superconducting gap. The inset in Fig.~\ref{edge-JDE}(b) shows the dependence of $\tilde{\Delta}$ on the junction length. In the short junction limit, $L/\xi\ll1$, $\tilde{\Delta}$ approaches $\Delta$, and the Andreev levels contributed by the two edges exhibit minimal differences. Thus, to capture the asymmetric edge states, we anticipate the junction length to be on the order of the coherence length ($L\sim\xi$). When an external magnetic field $B$ is applied to the junction area $S$, the Josephson potential of the system, can then be approximated as
\begin{equation}
V(\phi,\phi_B)=-\Delta_1\left|\cos\left(\frac{\phi_1}{2}\right)\right|-\Delta_2\left|\cos\left(\frac{\phi_2}{2}\right)\right|
\label{edge-pot}
\end{equation}
where $\Delta_{1(2)}$ indicates the eigenenergies of Josephson potential at $\phi=0$ for the two edges. The phase differences $\phi_{1(2)}$ are related by the magnetic flux across the junction, $\phi_2-\phi_1=\phi_B$, where $\phi_B=2\pi\phi_e/\phi_0$ with $\phi_e=BS$ the magnetic flux and, $\phi_0=h/2e$ the magnetic flux quantum. Note that asymmetric edge states acquire distinct additional phases due to the external magnetic flux, but these can be transformed into the form given in Eq.~\eqref{edge-pot} through variable substitution and have no effect on the diode efficiency.

Fig.~\ref{edge-JDE}(b) shows how the Josephson diode efficiency $\eta=(I_{c+}-|I_{c-}|)/(I_{c+}+|I_{c-}|)$ varies with the magnetic field. At $\phi_B=\pi$, the JDE disappears with $\eta=0$ because the Josephson potential in Eq.~\eqref{edge-pot} preserves the symmetry $V(\phi)=V(\pi-\phi)$. Furthermore, the relationship $\eta(\phi_B)=-\eta(2\pi-\phi_B)$ is evident due to the equivalence between the Josephson potential Eq.~\eqref{edge-pot} at $\phi_B=2\pi-\phi'_B$ and $\phi_B=-\phi'_B$ with a phase shift $\phi\to\phi-\pi$. This scenario effectively reverses the direction of the magnetic flux, leading to interference between the two edge states that results in opposite critical currents and a reversal of diode efficiency. Notably, it is the asymmetric edge that breaks the inversion symmetry in 1T'-WTe$_{2}$-based Josephson junctions and leads to the JDE. We subsequently analyze the diode efficiency as a function of different edge configurations while fixing the magnetic flux $\phi_B$ at the maximal diode efficiency point in Fig.~\ref{edge-JDE}(b). We retain the atomic configuration $\mathrm{Te_{B}}$ at the lower boundary and vary the upper boundary configuration by altering the upper limit $W+\delta d$ with $\delta d$ (schematically shown in the upper inset in Fig.~\ref{edge-JDE}(c)). The calculated diode efficiency is shown in Fig.~\ref{edge-JDE}(c), clearly indicating that no JDE occurs when the two edges are related by inversion symmetry (more results are shown in the lower inset in Fig.~\ref{edge-JDE}(c)), e.g., $\mathrm{W_A}$ and $\mathrm{W_B}$ edges, $\mathrm{Te_A}$ and $\mathrm{Te_B}$ edges. Additionally, the JDE polarity reverses if the mismatch in Fermi velocity between the two edges is reversed (depicted by the red and yellow lines in Fig.~\ref{edge-JDE}(c)) \cite{Li2024,Nikodem2024}. Utilizing the toy potential Eq.~\eqref{edge-pot} and defining the asymmetric parameter $r\equiv(\Delta_1-\Delta_2)/(\Delta_1+\Delta_2)$, we calculate the phase diagram of diode efficiency as a function of $r$ and magnetic flux $\phi_B$, illustrated in Fig.~\ref{edge-JDE}(d). It is evident that diode efficiency is an odd function of both $r$ and $\phi_B$ at $r=0$ and $\phi_B=\pi$, which is consistent with the numerical results. Note that, the Josephson potential form in Eq.~\eqref{edge-pot} cannot accurately describe the Josephson potential when the junction violates the short junction limit, $L/\xi\gg1$; more results are shown in Appendix B. The maximal diode efficiency is limited to about $30\%$, consistent with similar devices with asymmetric edge states \cite{Nikodem2024}.

\begin{figure}[!htbp]
	\centering
	\includegraphics[width=1\columnwidth]{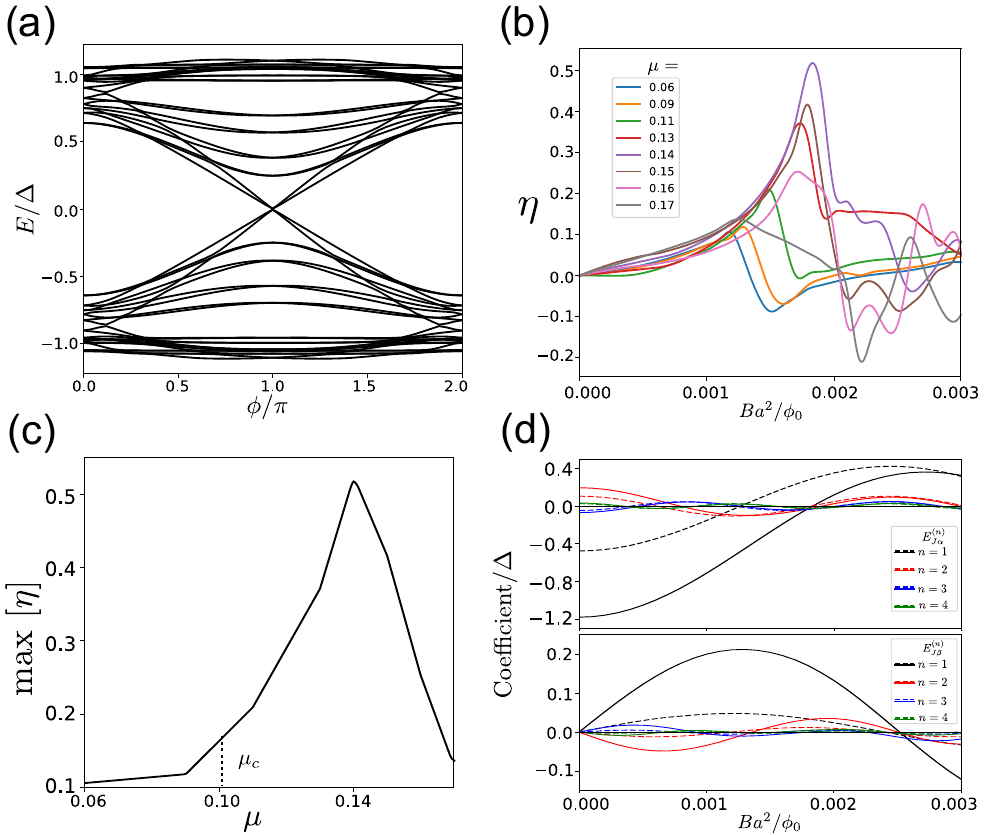}
	\caption{ (a) Andreev levels of the junction with a chemical potential in the bulk and slightly larger than $\mu_c$, $\mu=0.13$. (b) Diode efficiency changes with magnetic flux at different chemical potentials, changing from the gap ($\mu<\mu_c$) to the bulk state ($\mu>\mu_c$). (c) The maximal diode efficiency changes with chemical potential, extracted from (b), $\mu_c$ is the position of the bottom of the conduction band. (d) The Fourier coefficients as a function of magnetic flux with the chemical potential at $\mu=0.06$ (the dashed lines) and $\mu=0.14$ (the solid lines); the upper panel shows the coefficient of the cosine term; the lower panel shows the coefficient of the sine term.
  }
\label{BuEd}
\end{figure}

Having established the JDE in the 1T'-WTe$_2$ edge interferometer, we now show how to further enhance the diode efficiency. When the chemical potential is situated in the gap ($\mu\lesssim \mu_c$), the Josephson potential is primarily influenced by edge states. However, as the chemical potential is adjusted into the bulk states (as shown in Fig.\ref{band}(b)), contributions from both edge and bulk states significantly impact the potential. The Andreev levels of the system, when the chemical potential is within the bulk states, are calculated and depicted in Fig.\ref{BuEd}(a), clearly illustrating the bulk state's contribution to the Josephson potential. We then evaluate variations in diode efficiency with respect to magnetic flux for different chemical potentials, as shown in Fig.\ref{BuEd}(b). Due to the fluctuating mismatch in Fermi velocity among the edge states with varying chemical potential (shown in Fig.~\ref{band}(a)), the effective area encircled by these states is not constant. Consequently, the magnetic field for which the effective flux $\phi_B$ approaches $\pi$ is also different, as shown in Fig.~\ref{BuEd}(b). But it remains the function of the magnetic field when the chemical potential in the gap, $\mu\lesssim \mu_c$ (blue and yellow lines in Fig.~\ref{BuEd}(b)). As the Fermi surface is further regulated to the bulk states, the diode efficiency loses its odd symmetry for specific magnetic fields due to Josephson potential from bulk states (Fig.~\ref{BuEd}(b)). The maximal diode efficiency can attain up to $53\%$ and then decrease with the chemical potential (Fig.~\ref{BuEd}(c)). Notably, adjusting the chemical potential simultaneously modulates the Fermi velocity mismatch between the two edges (Fig.~\ref{band}(a)) and the count of transport channels and their Fermi velocities of bulk states. Additionally, as the chemical potential approaches the bottom of the conducting band, the junction no longer conforms to the short junction limit due to low Fermi velocities in the edge and bulk states. Generally, the Josephson potential in this scenario can be expressed using a Fourier series as
\begin{equation}
\begin{aligned}
V'(\phi,\phi_B)=\sum_nE_{J\alpha}^{(n)}(\phi_B)\cos(n\phi)+E_{J\beta}^{(n)}(\phi_B)\sin(n\phi),
    \label{pot-eb}
\end{aligned}
\end{equation}
where $E_{J\alpha(\beta)}^{(n)}$ are the coefficients of the cosine and sine terms, respectively, and $n$ labels the Fourier order. With the computed Josephson potential, we analyze these coefficients as a function of magnetic flux with the chemical potential $\mu=0.06$ (indicating low diode efficiency with chemical potential in the gap) and $\mu=0.14$ (achieving maximal diode efficiency with chemical potential in bulk), as shown in Fig.~\ref{BuEd}(d). The coefficients exhibit periodic variations with magnetic flux when the chemical potential is within the bulk gap (the dashed lines in Fig.~\ref{BuEd}(d)), consistent with the periodicity of the diode efficiency in Fig.~\ref{BuEd}(b) and Fig.~\ref{edge-JDE}(b). Conversely, they do not vary periodically with the magnetic flux when the chemical potential is in the bulk states. Additionally, Fig.~\ref{BuEd}(d) clearly depicts the enhancement of both $E_{J\alpha}^{(n)}$ and $E_{J\beta}^{(n)}$, attributed to bulk state contributions and increased asymmetric edge dispersions, which is the key point for the generation of large diode efficiency. In particular, the Josephson potential contributed by the bulk states relies not only on the chemical potential but also on the width ($W$) of the junction. However, asymmetric edges do not depend on the width of the junction. Thus, the position of the chemical potential that achieves the maximum diode efficiency is related to the width of the junction. Furthermore, this enhancement of diode efficiency via bulk states alongside substantial asymmetric edge dispersions is a universal finding, supported by a toy potential (Appendix C).

\section {Discussion and summary}

In this study, we have systematically investigated the Josephson diode effect in a junction based on monolayer 1T'-WTe$_2$. Our central results demonstrate that the intrinsic asymmetric edge terminations of WTe$_2$ nanoribbons provide a powerful resource for engineering a JDE.  While the effect driven by topological edge states alone is significant, we have uncovered a crucial mechanism to further enhance the diode efficiency: tuning the chemical potential to allow for the participation of bulk states. This interplay boosts the maximum efficiency to more than $50\%$, a highly competitive value.

While the TB model in the main text captures the essential physics and is widely used for 1T'-WTe$_2$ \cite{Ok2019,Hsu2020,Nandy2022}, it has been shown that a more detailed model including additional spin-orbit coupling terms might be needed for a precise quantitative comparison with experiments \cite{Lau2019}. Nevertheless, the core mechanism—the existence of inequivalent edge states—is a robust feature of WTe$_2$, and we expect our main conclusions to hold. Excitingly, with recent experimental advances in realizing gate-induced superconductivity in WTe$_2$ flakes \cite{Sajadi2018}, our proposed device could potentially be implemented in an all-in-one material platform using local gating techniques, avoiding complex fabrication of hybrid structures. 

Moreover, the edges of the samples in experiments on 2D monolayer materials may be irregular, i.e., they have cracks, steps, or bumps \cite{Wu2018}. Hence, such samples lose translational invariance along their boundaries, resulting in a variable Fermi velocity mismatch between the two edges, which further influences the JDE within the system. However, we find that high diode efficiency is maintained when the Fermi level is in the bulk gap with the rough boundaries (Appendix D).

In conclusion, our work establishes a novel and practical route toward achieving a highly efficient and robust JDE. The gate-controlled technology can enhance the diode efficiency in an experiment. The principle of enhancing diode efficiency by admixing bulk transport channels is general and could be applied to other topological materials with asymmetric edge states \cite{Chen2018}, such as Ta${_2}$Pd${_3}$Te${_5}$ \cite{Li2024} and MnBi$_2$Te$_4$ \cite{Gao2024,Zhang2025}. These findings are not only of fundamental interest but also pave the way for the development of next-generation dissipationless quantum electronics.


\begin{acknowledgments}
We acknowledge useful discussions with Noah F. Q. Yuan, Shun Wang, and Cheng-Yu Yan. X. Liu acknowledges the support of the Innovation Program for Quantum Science and Technology (Grant No. 2021ZD0302700), the National Natural Science Foundation of China (NSFC) (Grant No. 12074133), the Shanghai Science and Technology Innovation Action Plan (Grant  No. 24LZ1400800) and the Project supported by Cultivation Project of Shanghai Research Center for Quantum Sciences (Grant No. LZPY2024).
\end{acknowledgments}

\begin{appendix}

\section{band structure of WTe$_2$}
\begin{figure}[!htbp]
	\centering
	\includegraphics[width=1\columnwidth]{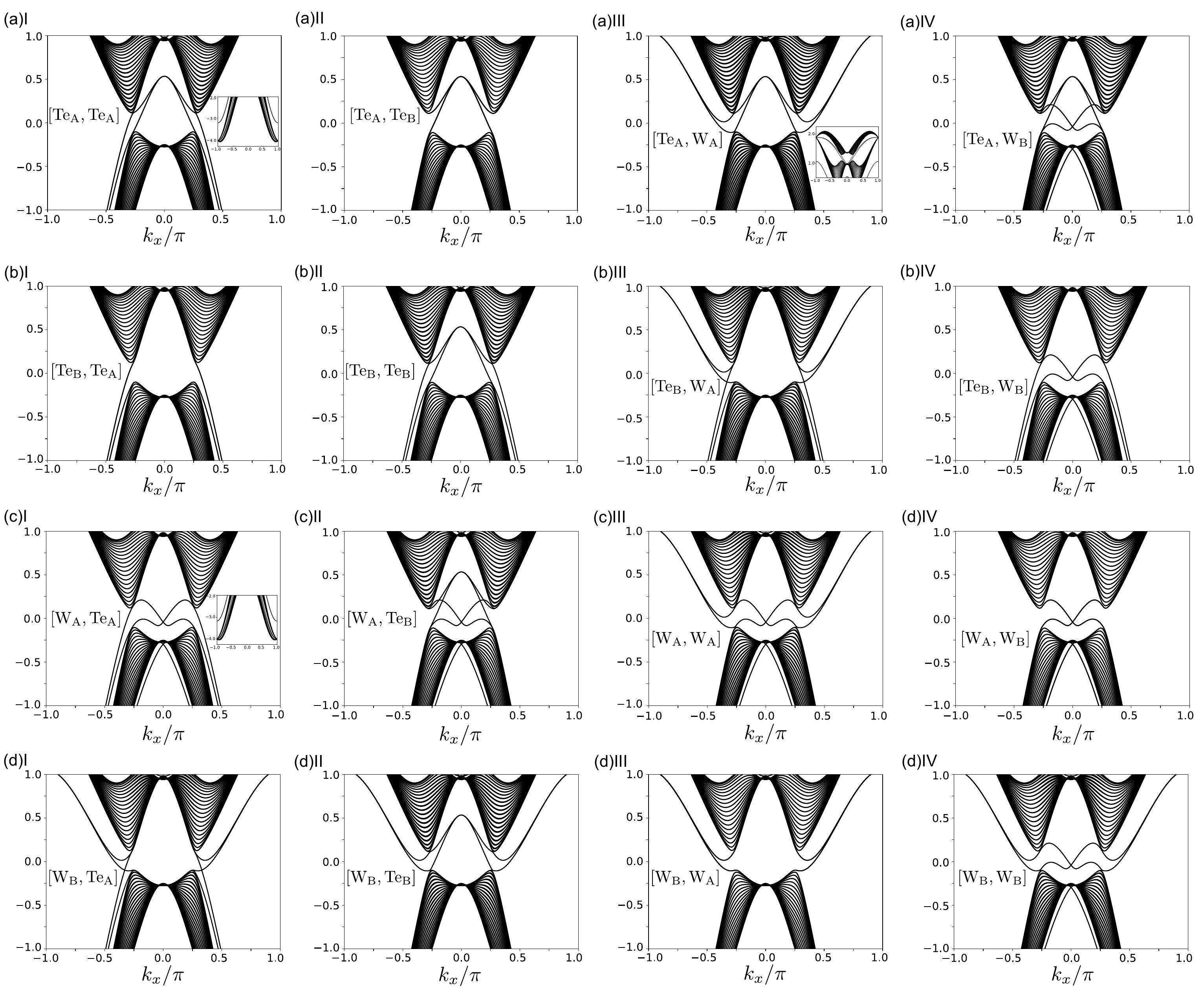}
	\caption{ With periodic condition in $x$ direction and width $W=11$. Band structures of all 16 terminal configurations. The inset shows the configuration of the terminations.
  }
\label{fu-band}
\end{figure}

Here, we demonstrate the band structure of the WTe$_2$ slab along the x-direction. There are possible $4\times4=16$ configurations of the slab along the x-direction in total. However, due to the inversion symmetry of the unit cell, there are only ten unique configurations. The redundancies can be noted as follows $[\mathrm{W_A,W_A}]=[\mathrm{W_B,W_B}]$, $[\mathrm{W_A,Te_A}]=[\mathrm{Te_B,W_B}]$, $[\mathrm{W_A,Te_B}]=[\mathrm{Te_A,W_B}]$, $[\mathrm{Te_A,W_A}]=[\mathrm{W_B,Te_B}]$, $[\mathrm{Te_A,Te_A}]=[\mathrm{Te_B,Te_B}]$, $[\mathrm{W_B,Te_A}]=[\mathrm{Te_B,W_A}]$. The complete band structure for all 16 terminal configurations is depicted in Fig.~\ref{fu-band}, where equivalent cases are clearly identified.

\begin{figure}[!htbp]
	\centering
	\includegraphics[width=1\columnwidth]{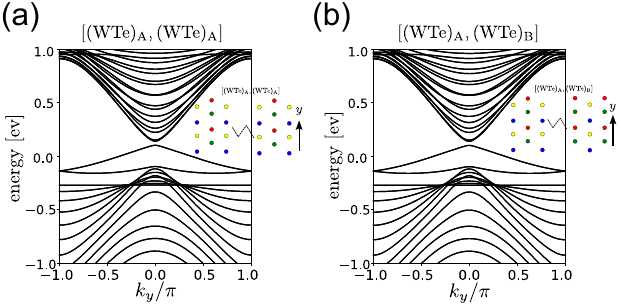}
	\caption{ Band structure with the periodic condition in the y-direction, open boundary condition in the x-direction with the length of $L=15$, with terminations $[\mathrm{(WTe)_A, (WTe)_A}]$ (a) and $[\mathrm{(WTe)_A, (WTe)_B}]$ (b). Insets show the structure for the terminations $\mathrm{[(WTe)_A, (WTe)_A]}$ and $\mathrm{[(WTe)_A, (WTe)_B]}$, the two edges of the termination $\mathrm{[(WTe)_A, (WTe)_A]}$ are related by the $M_x$ symmetry (the dashed green line).
  }
\label{band--y}
\end{figure}

With the periodic boundary condition in the $y$ direction, the band structures are shown in Fig.~\ref{band--y}. There are possible $2\times2=4$ configurations of the slab along the $y$-direction in total. However, due to the inversion symmetry of the unit cell and the mirror symmetry $M_y$ of the slab along the $y$ direction, all the band structures are equivalent. Fig.~\ref{band--y} shows the band structures with terminations $[\mathrm{(WTe)_A, (WTe)_A}]$ and $[\mathrm{(WTe)_A, (WTe)_B}]$, displaying identical characteristics.

\section{Diode efficiency changes with magnetic flux with the chemical potential in the gap}
\begin{figure}[!htbp]
	\centering
	\includegraphics[width=1\columnwidth]{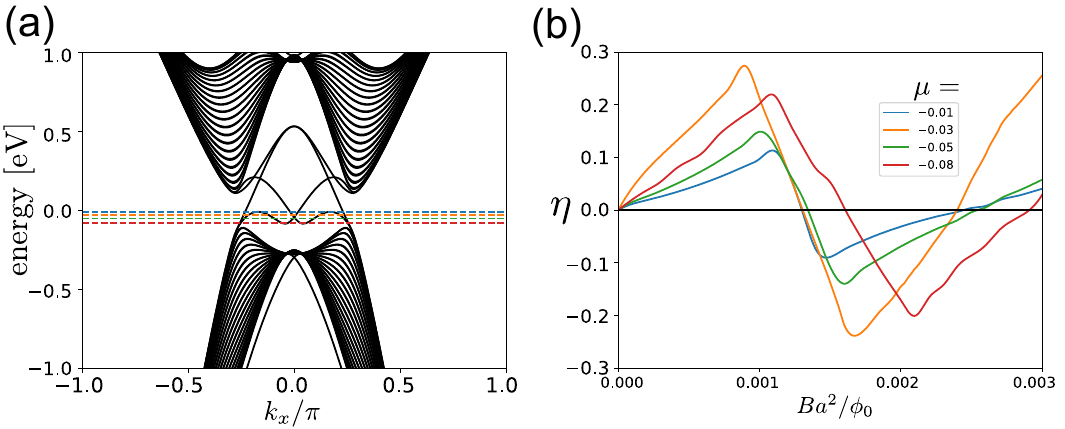}
	\caption{ Diode efficiency as a function of magnetic flux with the chemical potential in the gap. (a) The band structure and (b) the diode efficiency as a function of magnetic flux with different chemical potentials in (a).
  }
\label{eta-gap-2}
\end{figure}

Here, we present additional results on diode efficiency as a function of magnetic flux, with the chemical potential located within the gap. Adjusting the chemical potential alters the mismatch in the Fermi velocity of the edge states, consequently modifying the coherence length of these states. When the junction exceeds the short junction limit $L/\xi\gg1$, the Josephson potential cannot be accurately approximated using the form in Eq.~\eqref{edge-pot}. Instead, it can be represented by the Fourier series in Eq.\eqref{pot-eb}. Fig~\ref{eta-gap-2}(a) illustrates the band structure alongside the chemical potential configuration where the junction length $L$ notably exceeds the coherence length $\xi$. Fig.~\ref{eta-gap-2}(b) depicts the diode efficiency as a function of magnetic flux with the chemical potential marked in (a). The maximum diode efficiency reaches approximately $30\%$ \cite{Pal2022,Guo2024}.

\section{Diode effect with toy Josephson potential}
In the 2DTI based Josephson junction, the asymmetric edges and magnetic flux play a crucial role in generating the JDE, with diode efficiency primarily constrained by the asymmetry of the edges alone. However, when Josephson currents contributed by bulk states are considered, diode efficiency can be significantly enhanced. To further establish the universality of this conclusion, we consider the ideal scenario in which the junction adheres to the short junction limit for edge states, and bulk states conform to the Fraunhofer diffraction pattern averaged over magnetic flux. This differs from the main text scenario, which violates the short junction limit and includes few bulk channels. The toy model for the Josephson potential can be formulated as

\begin{figure}[!htbp]
\centering
\includegraphics[width=1\columnwidth]{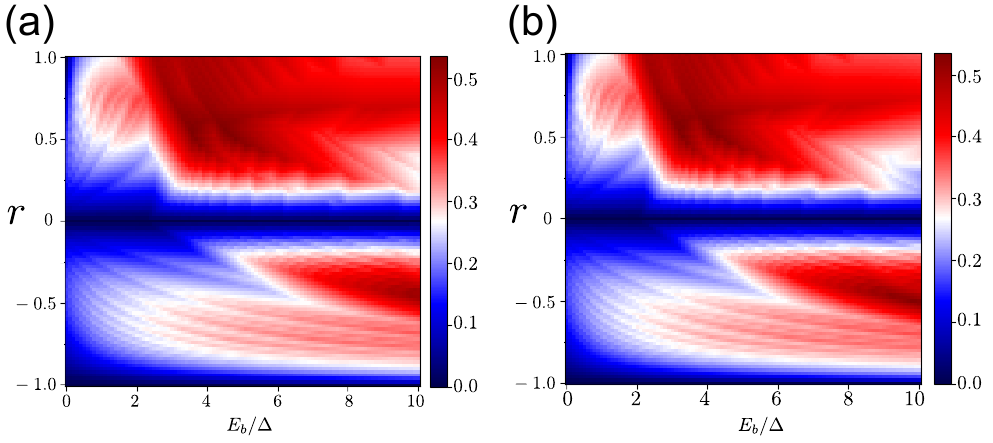}
\caption{ The maximal diode efficiency changes with $r$ and $E_b$ with the toy potential Eq.~\eqref{toy-po-as} with $k_1\approx k_2\approx k_3=1$. (a) and (b) are the results with $\Delta_1=0.85\Delta,0.95\Delta$, respectively.
}
\label{eta-toy-m}
\end{figure}

\begin{equation}
\begin{aligned}
    V'(\phi,\phi_B)&=-\Delta_1\left|\cos\left(\frac{2\phi-k_1\phi_B}{4}\right)\right|-\Delta_2\left|\cos\left(\frac{2\phi+k_2\phi_B}{4}\right)\right|\\
    &-E_b\cos\phi\frac{\sin (k_3\phi_B)}{(k_3\phi_B)},
\label{toy-po-as}
\end{aligned}
\end{equation}
where $E_b>0$ the amplitude of the Josephson potential contributed by bulk states, and can be tuned by the chemical potential. $k_{1,2,3}$ characterizes the phase of the edge states and bulk states due to magnetic flux. Note that the asymmetric edges with different Fermi velocities have different decay length, so the Josephson potential contributed by the two edges will gain different additional phases in the presence of magnetic flux ($k_1$ is not exactly equal to $k_2$), which can not be transformed into the form of Eq.~\eqref{edge-pot} in the presence of Josephson potential contributed by bulk states. The area enclosed by the edge states will be slightly smaller than the junction area due to the finite decay length of edge states. The magnetic flux enclosed by the junction area is not strictly equal to the additional phase differences of the two edge states ($(k_1+k_2)/2$ is not exactly equal to $k_3$). However, in experiments, we expect the junction width to be significantly larger than the decay length of edge states. Therefore, we approximate $k_1\approx k_2\approx k_3=1$ to explore the enhancement of diode efficiency with bulk states. By keeping $\Delta_1$ fixed and defining asymmetry parameter $r\equiv(\Delta_1-\Delta_2)/(\Delta_1+\Delta_2)$, we calculate the maximal diode efficiency as a function of $r$ and $E_b$, shown in Fig.~\ref{eta-toy-m}. It clearly demonstrates that bulk state contributions significantly enhance the maximal diode efficiency with large asymmetric edges.

\section{JDE with disordered edges}
\begin{figure}[!htbp]
	\centering
	\includegraphics[width=1\columnwidth]{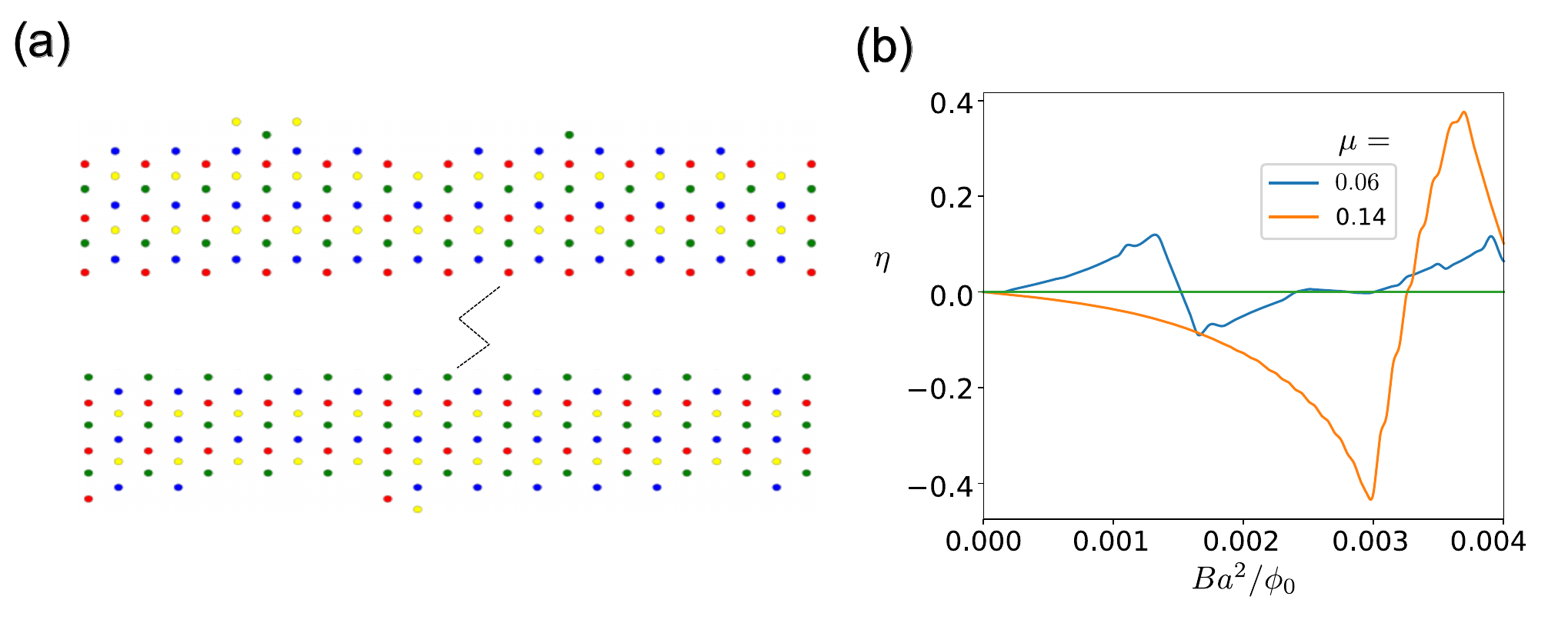}
	\caption{(a) Geometries of the system with rough edges, corresponding to adding or removing two sites along the boundary randomly. (b) The diode efficiency as a function of magnetic flux with the rough edges shown in (a).
  }
\label{dis-jde}
\end{figure}

Here, we model a rough edge by the boundary following a random step $\delta y$ appearing at each site along an initially translationally invariant boundary, where $\delta y$ is randomly and uniformly distributed within the range of $[-\delta p,\delta p]$ and exhibits a zero mean. The parameter $\delta p$ thus quantifies the degree of boundary irregularity. Fig.~\ref{dis-jde}(a) illustrates an irregular system, which permits the addition or removal of up to one site along the boundary. This value suffices for experimentation as an atomically flat interface can be realized \cite{Zhu2023,Li2024}. With the rough edge system, we calculate the variations in Josephson diode efficiency with magnetic flux when the Fermi level is situated within the bulk gap and bulk states, as depicted in Fig.~\ref{dis-jde}(b). The rough edge exerts minimal impact on diode efficiency when the Fermi level resides in the bulk gap. However, it significantly influences the performance when the Fermi level is positioned at the bottom of the conduction band due to the low Fermi velocities of edge and bulk states. Nonetheless, a high diode efficiency is maintained when the Fermi level is in the bulk gap.

\end{appendix}

%

\end{document}